\documentclass[aps,pre,preprint,superscriptaddress]{revtex4-1}

\usepackage{graphicx}

\begin{document}

\draft
\title{Comparison of Domain Nucleation Mechanisms in a Minimal Model of Shoot Apical Meristem}


\author{Dorjsuren Battogtokh}
\email[]{dbattogt@vt.edu, tyson@vt.edu}
\affiliation{The Institute of Physics and Technology, Mongolian Academy of Sciences, Ulaanbaatar 51, Mongolia}
\affiliation{ Department of Biological Sciences, Virginia Polytechnic and State University, Blacksburg, Virginia 24061, USA}
\author{John J. $\rm {Tyson^{*,}}$}
\affiliation{ Department of Biological Sciences, Virginia Polytechnic and State University, Blacksburg, Virginia 24061, USA}
\email[]{tyson@vt.edu}



\date{\today}

\begin{abstract}
Existing mathematical models of the shoot apical meristem (SAM) explain nucleation and confinement of a stem cell domain by Turing's mechanism, assuming that the diffusion coefficients of the activator (WUSCHEL) and inhibitor (CLAVATA) are significantly different. As there is no evidence for this assumption of differential diffusivity, we recently proposed a new mechanism based on a “bistable switch” model of the SAM. Here we study the bistable-switch mechanism in detail, demonstrating that it can be understood as localized switches of WUSHEL activity in individual cells driven by a non-uniform field of a peptide hormone. By comparing domain formation by Turing and bistable-switch mechanisms on a cell network, we show that the latter does not require the assumptions needed by the former,  which are not supported by biological evidences.
\end{abstract}


\maketitle

\section{Introduction}
 The stem cells residing in the shoot apical meristem (SAM) give rise to above ground tissues \cite{laux}. Hence, maintenance of stem cell niches is of central importance to plant growth  \cite{meyer0,meyer}. A negative feedback between the proteins WUSCHEL (WUS - a homeodomain transcription factor) and CLAVATA (CLV - a receptor kinase) is at the core of the signaling pathway controlling the central domain –- the reservoir of stem cells  \cite{laux}. The cell-to-cell communications orchestrated by the CLV-WUS network in the SAM are not fully understood and a detailed quanitative model of SAM can be insightful. In particular,  the underlying mechanism of pattern formation is crucial in understanding how the size, location and stability of the central domain is controlled in the SAM.
  
Recently, reaction-diffusion models of SAM  have been studied, using Turing instability \cite{meyer1,simon,fujita,nikolaev} as the mechanism of domain nucleation and confinement. Turing instability is the most well known mechanism of pattern formation in dissipative systems \cite{turing}, with the critical condition that the diffusion length of an inhibitor significantly exceeds the diffusion length of an activator \cite{mein,murray}. Under this condition, a periodic pattern emerges in monostable system, at a certain critical wavenumber, from small non-uniform perturbations of uniform solutions. For modeling stem cell nucleation in the SAM, the Turing mechanism requires that the diffusion coefficient of CLV (inhibitor: the complex of membrane receptor kinase CLV1-CLV2 and its ligand CLV3) significantly exceeds that of WUS (activator). Accordingly, an area under a Turing pattern, where WUS concentration exceeds the steady--state level can be identified as a stem cell domain; whereas the areas where WUS levels are below the steady state level can be identified with other SAM zones. The selection of the critical wavenumber ensures a fixed domain size, while enforcing of a spatial heterogeniety of 
a certain parameter confines the domain at a given location \cite{simon}.  At present, the diffusion coefficients of CLV and WUS have not been measured; therefore, there is no clear experimental evidence on whether the Turing condition of differential diffusivity is applicable within the WUS and CLV expression zones of the SAM.

Existing models of the SAM regulation involve positive and negative feedback loops that can generate not only Turing patterns but also alternative stable steady states (bistability) in a certain range of parameter values  \cite{meyer0,fujita,nikolaev}. For a bistable-model of SAM, an area
where WUS's distribution is near to an upper steady state represents a stem-cell domain; whereas, the areas near to a lower steady-state represent other zones of SAM. A sufficiently strong perturbation can nucleate of an upper WUS domain from the lower state, and because the upper state is a stable solution of the reaction system, the domain can be confined at a given location if the fronts connecting upper and lower domains are motionless.  Recently bistable reaction-diffusion models have been studied to simulate experimental data on cytokinin-controlled domain confinement in SAM \cite{meyer}. 

In our previous work \cite{bat0}, a mechanism different from Turing instability was proposed for pattern formation in a minimal, bistable model of SAM.  In the present  work, we study in detail how a spatially nonuniform field of a peptide-hormone synthesized in the system drives domain nucleation at the onset of bistability. Our goal is to
compare domain formations by Turing and bistable-switch mechanisms side-by-side, on an array of cells, as well as  
on a polygonal  cell-network. Obviously, a model that not only explains biological facts correctly, but also makes a valid predictions should be prefferred for modeling of the SAM. Because of the importance of spatial regulations  in the SAM, understanding the underlying mechansim of domain formation is essential for selecting the right model.  

This work is organized as follows. In section II we introduce a minimal, three-variable model of SAM. In section III, we study a bifurcation diagram,  stationary solutions, and stability of uniform solutions of the activator-inhibitor core of the model. Section IV illustrates in detail the domain nucleation mechanisms in the two-variable, activator-inhibitor model. In section IV, it is also shown that in the minimal model, the fast diffusive field of the agent plays the role of the bifurcation parameter for domain nucleation. In Section V we simulate  Turing  and ``bistable switch'' models in on a two-dimensional cell-network - a model of a longitudinal cross-section of SAM.  We simulate  well-known patterns displayed by the central zone in Section V. The last section is devoted to discussion. 


\section{A Minimal Model of the SAM}
\begin{figure}
\vspace{-0.1in}
\includegraphics[width=1.\linewidth,height=1.\linewidth,scale=0.6]{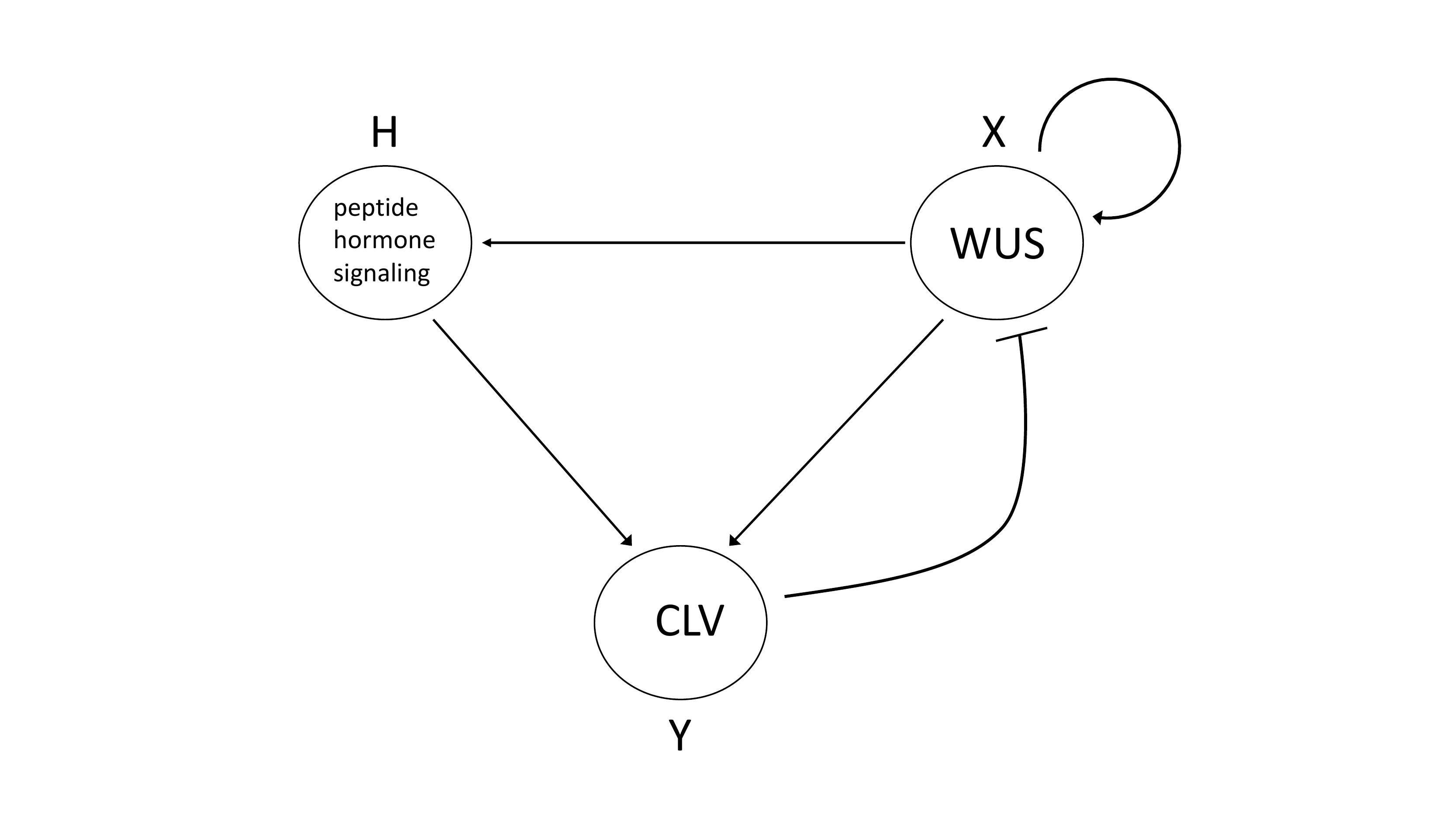} 
\caption{\label{fig1} Wiring diagram of a minimal model.}
\end{figure}
In Fig. \ref{fig1} we present a wiring diagram of a three-variable model of SAM \cite{bat0}, which describes the interaction between the key proteins involved in the stem cell regulation of a model plant  {\em Arabidopsis}. We note that Nikolaev {\em et al} \cite{nikolaev}  earlier introduced a three-variable, minimal model of SAM, which differs from Fig. \ref{fig1} with regards: 1) $X$ is not self-enhanced, but activated by $H$, and 2) $Y$ is not directly activated by $X$. We adopted these two changes from Fujita {\em et al} \cite{fujita}, which lead to a simpler model. However, Fujita {\em et al}  studied domain formation by Turing's mechanism. On the contrary, we study  domain formation in the wiring diagram of Fig. \ref{fig1} by the bistable-switch mechanism  \cite{bat0}. We believe that the mechanism of domain formation in Nikolaev {\em et al} has the same origin as our bistable-switch mecahnism.

\begin{figure}
\vspace{-0.1in}
\includegraphics[width=0.7\linewidth,height=0.4\linewidth,scale=0.5]{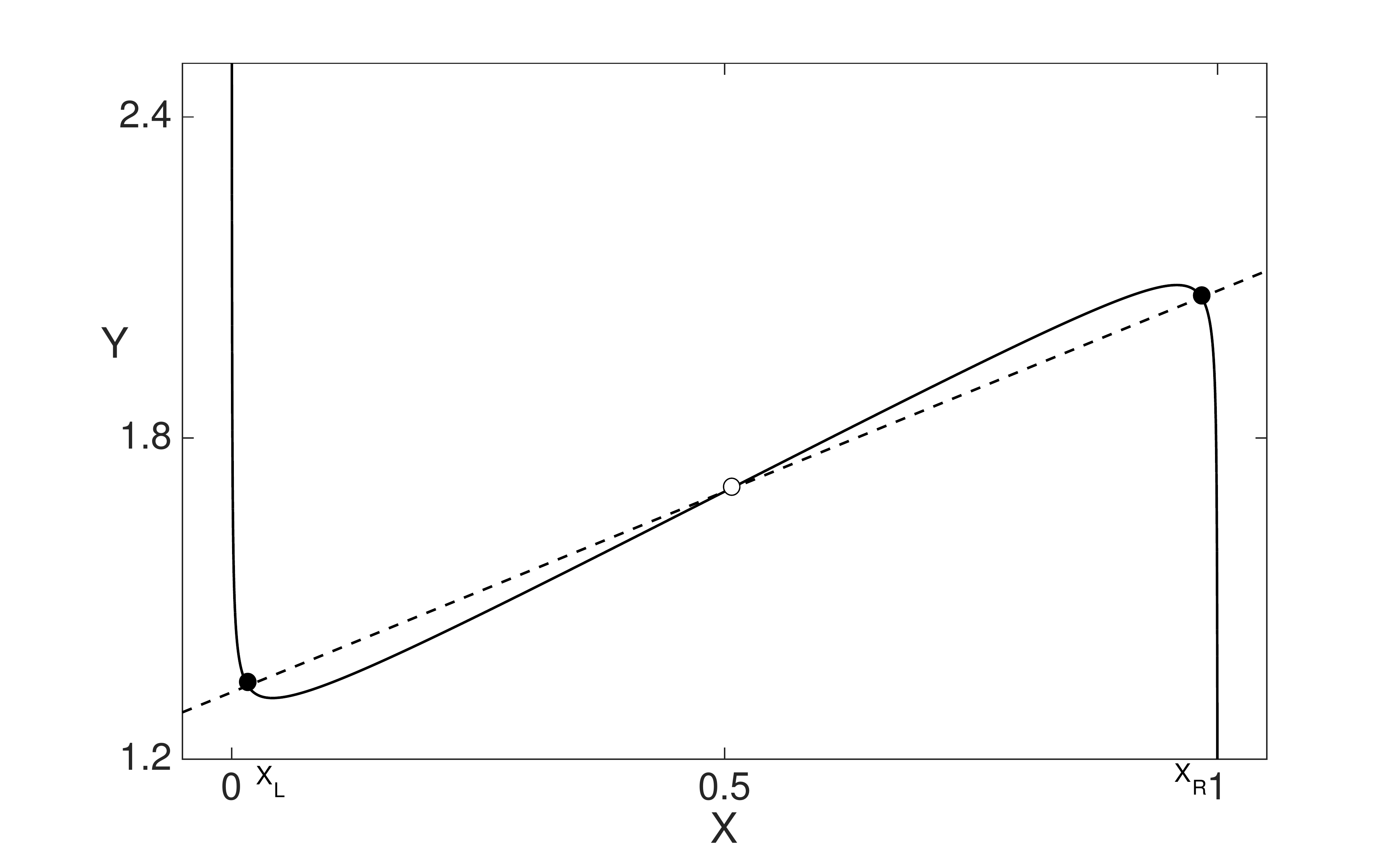} 
\caption{\label{fig2} Nullclines in Eq. (4). Parameters are: $\gamma=0.75$, $\alpha=1.2$, $\sigma^2=0.001$, and $\beta=1.325$. }
\end{figure}

Converting the wiring diagram in Fig. 1 into a mathematical model of discrete cells \cite{nikolaev,fujita,bat0,kura},  we obtain the following set of partial differential equations, 
\begin{eqnarray}
\frac{\partial  X_i}{\partial t} & = & D_X (X_{i-1}-2 X_i + X_{i+1}) -X_i +\Phi_{\sigma}(\alpha+ X_i -Y_i), \nonumber\\
\frac{\partial  Y_i}{\partial t} & = & D_Y (Y_{i-1}-2 Y_i + Y_{i+1}) +\frac{1}{\epsilon_Y}(\hat Y_0 +\mu H(x_i)  + \gamma X_i -Y_i),  \nonumber\\
\frac{\partial  H}{\partial t} & = & D_H \Delta H+ \frac{1}{\epsilon_H}(\sum\limits_{j=1}^N X_j \delta (x-x_j) - H), 
\end{eqnarray}
where $X_i(t)$ denotes the concentration of the master protein $WUS$  in a cell $i$ ($1\leq i \leq N$) , $Y_i(t)$ describes the complex of $CLV3$ and $CLV1\&CLV2$ in a cell $i$, and $H$ describes a hypothetical peptide-hormone.  $D_X$, $D_Y$, and $D_H$ denote diffusion coefficients of WUS, CLV, and H, respectively. Previous mathematical models describing experimental data on SAM development suggested the existence of a hypothetical peptide-hormone, which was called {\em stemness factor} in Ref. \cite{simon}. $H(x,t)$ in Eq. (1) is the peptide-hormone's level in the SAM region. Greek symbols in Eq. (1) represent positive parameters and $N$ is the number of cells. $\hat Y_0$ is the basal expression level of the CLV complex. The  function $\Phi_{\sigma}$ describes a nonlinear sigmoidal regulation of $WUS$ expression \cite{meyer1}, 
\begin{eqnarray}
\Phi_{\sigma}(\xi) & = &\frac{1}{2}(1+\frac{\xi}{\sqrt{\sigma^2+\xi^2}}),\nonumber\\
\xi & = & \alpha+X_i-Y_i,
\end{eqnarray}
Depending on whether $\xi>0$ or $\xi<0$, $\Phi_{\sigma}$ can switch  $X$'s expression between off and on states. The smaller is $\sigma$, the stiffer is the switch. 
\begin{figure}
\vspace{-0.1in}
\includegraphics[width=1.\linewidth,height=0.5\linewidth,scale=0.4]{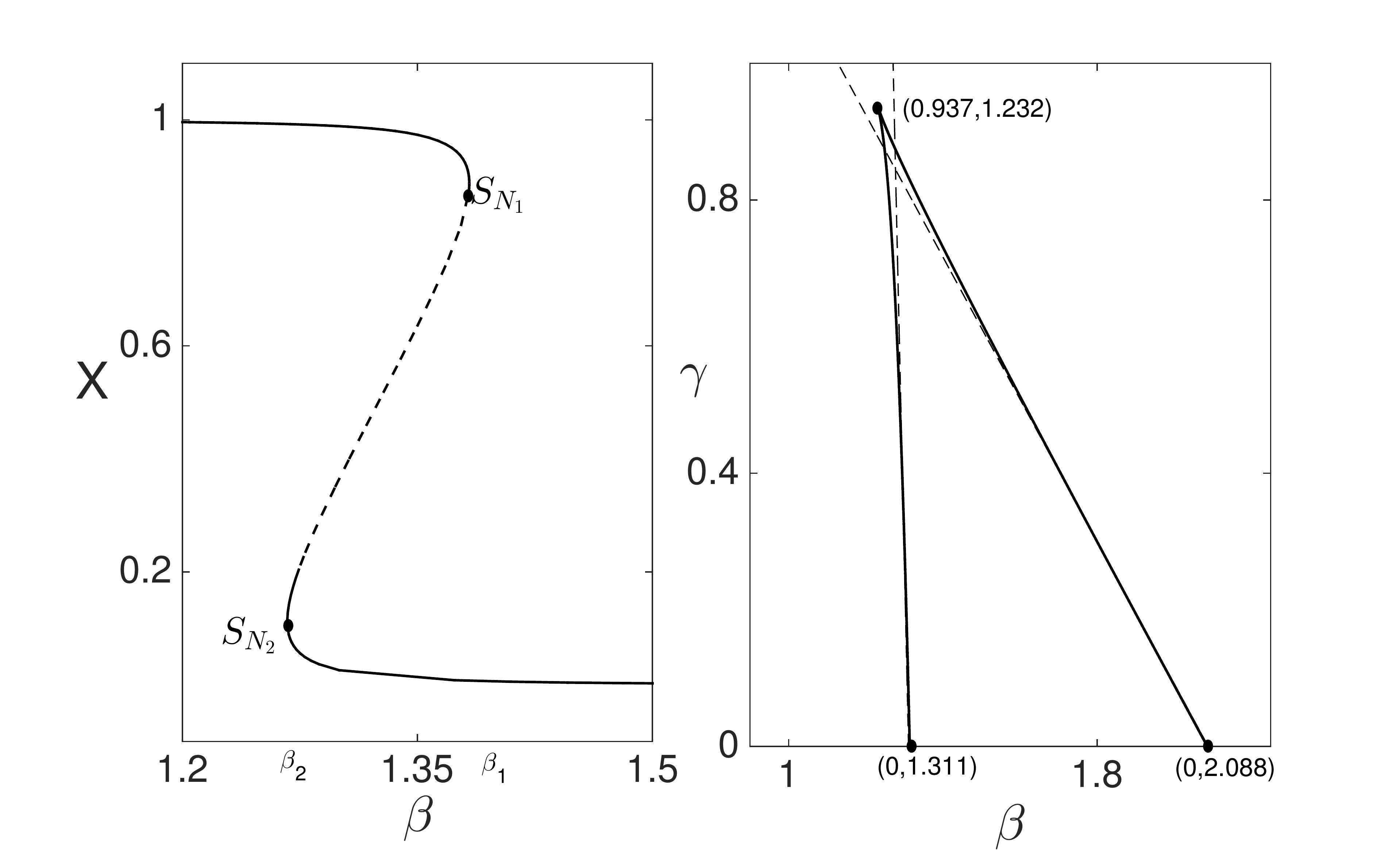} 
\caption{\label{fig2} Bistability in the reaction model of Eq. (4). A) Saddle-node bifurcations of $X$ for $\beta$ used as the bifurcation parameter. B) Cusp bifurcation. Dashed lines are obtained by the formula  $\gamma=\frac{Y_{L,R}}{X_{L,R}}-\frac{\beta}{X_{L,R}}$. Parameters are: $\gamma=0.75$, $\alpha=1.2$, $\sigma^2=0.001$, $\beta=1.275$, and $\epsilon_Y=1$. }
\end{figure}

\section{A two variable model}
Let us  assume  $\epsilon_H<<1$ and $D_H>>1$, which imply that $H$ is a fast-diffusive variable. Reaction-diffusion models coupled through a fast diffusive field have been studied previously for models with oscillatory dynamics \cite{kura,kura1,battphysA,chimera}. In the limit of $\epsilon_H<<1$ and $\epsilon_H<<\epsilon_Y$, the last equation of Eq. (1) can be approximated by 
\begin{equation}
\hat H(x,t) \sim \int_0^L \mathrm{e}^{-\frac{ |x-x'|}{\sqrt{\epsilon_H D_H}}}X(x',t)\,\mathrm{d}x'.
\end{equation}
For $\sqrt{\epsilon_H D_H} >> L$, where $L$ is the system size, $\hat H(x,t)$ can be replaced by the global coupling function of $X$, $\hat H_g \approx {\overline X}$. Let us consider the case of $\overline X =const$, and introduce a constant, $\beta= \hat Y_0+ \mu \overline X$. Then Eq. (1) can be reduced to a two-variable model,
\begin{eqnarray}
\frac{\partial  X_i}{\partial t} & = &D_X  (X_{i-1}-2 X_i + X_{i+1})  -X_i +\Phi_{\sigma}(\alpha+ X_i-Y_i), \nonumber\\
 \epsilon_Y \frac{\partial  Y_i}{\partial t} & = & D_Y  (Y_{i-1}-2 Y_i + Y_{i+1}) +(\beta  + \gamma X_i -Y_i).
\end{eqnarray}
We note that Fujita {\em et al} Ref. \cite{fujita} first introduced Eq. (4) for a different sigmoidal function  $\Phi$, in the case of $\beta=0$ and $\epsilon_Y=1$, as a model of an ``activator-inhibitor''  interaction between WUS and CLV3 proteins. In Ref. \cite{fujita}, Turing patterns have been simulated on dynamic cell networks, for modeling experimental data on SAM, using different modifications of  Eq. (4).  


\begin{figure}
\vspace{-0.1in}
\includegraphics[width=1.\linewidth,height=0.5\linewidth,scale=0.4]{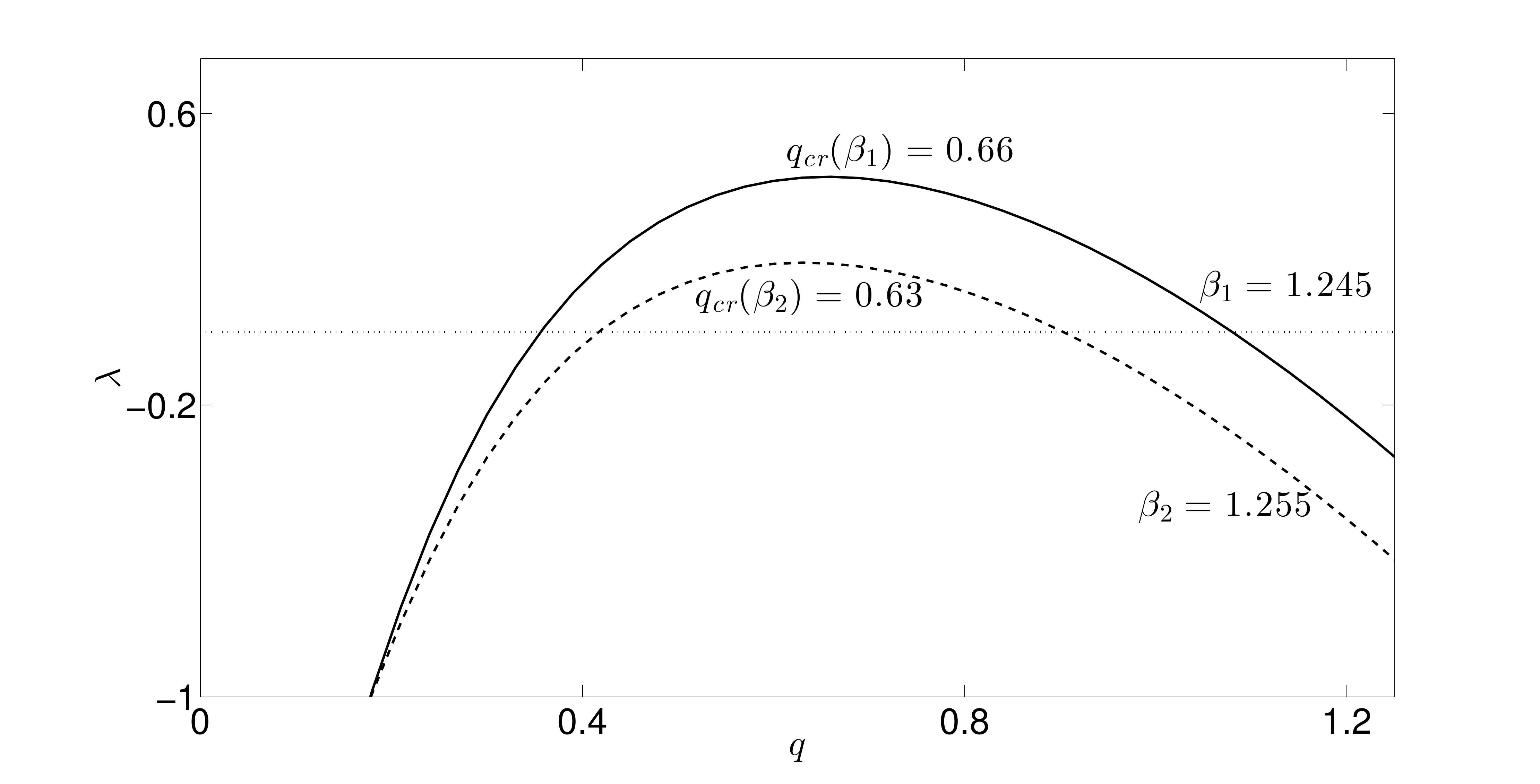} 
\caption{\label{fig3} Turing instability of uniform solutions. Parameters are:  $\alpha=1.2$, $\gamma=1.2$,   $\sigma^2=0.006$, $\epsilon_Y=0.1$, $D_X=1$, and $D_Y=10$. }
\end{figure}

\subsection{Bifurcation diagram and stationary solutions}
Since we study only the reaction part of Eq. (4)  in the following subsections, we remove the index $i$ of the variables.
Fig. \ref{fig2}A shows a bifurcation diagram of Eq. (4), where  $\beta$ is used as the principal bifurcation parameter. There are two saddle-node points $S_{N_1}$ and $S_{N_2}$ in Fig. \ref{fig2}A, connected by unstable steady states. For  $S_{N_2} < \beta < S_{N_1}$  Eq. (4) displays bistability.

The $X$ and $Y$ nullclines are,
\begin{eqnarray}
Y & = & \alpha +X - \sigma  \frac{X-\frac{1}{2}}{\sqrt{X(1-X)}}, \nonumber\\
Y & = & \beta + \gamma X.
\end{eqnarray}
From the intersections of these nullclines(see Fig. 2), bistable steady states can be determined. By replacing $Y$ in Eq. (5) with $Y=\beta + \gamma X$, the stationary solutions of Eq. (4)  can also be found by solving
\begin{equation}
\alpha -\beta + (1-\gamma) X  =  \sigma \frac{X-\frac{1}{2}}{\sqrt{X(1-X)}}.
\end{equation}
Eq. (6) can have three solutions,  $0 <X_1<X_0 <X_2$, in a certain range of the parameter values. For the special case $\beta = \alpha + \frac{1}{2}(1-\gamma) > 0$, the solutions are $X_{0}=\frac{1}{2}$  and $X_{1,2}=\frac{1 \pm \sqrt{1-(\frac{\sigma}{\alpha-\beta})^2}}{2}$.

The maximum and minimum of the $X$ nullcline are given by,
\begin{eqnarray} 
X_{{R,L}}=\frac{1}{2} \pm \frac{1}{2} \sqrt{1-(4 \sigma^2)^ \frac{1}{3}}, \nonumber\\
Y_{{R,L}}=\alpha +\frac{1}{2} \pm \frac{1}{2} [1-(4 \sigma^2)^{\frac{1}{3}}]^ \frac{3}{2}.
\end{eqnarray}

The saddle-node bifurcation points $\beta_{1,2}$ (dashed lines in the left plot of Fig. 2) are given approximately by $\beta_{1,2} \approx Y_{R,L} - \gamma X_{R,L}$, with strict equality holding only for $\gamma=0$. 

\begin{figure}
\vspace{-0.1in}
\includegraphics[width=1.\linewidth,height=0.5\linewidth,scale=0.4]{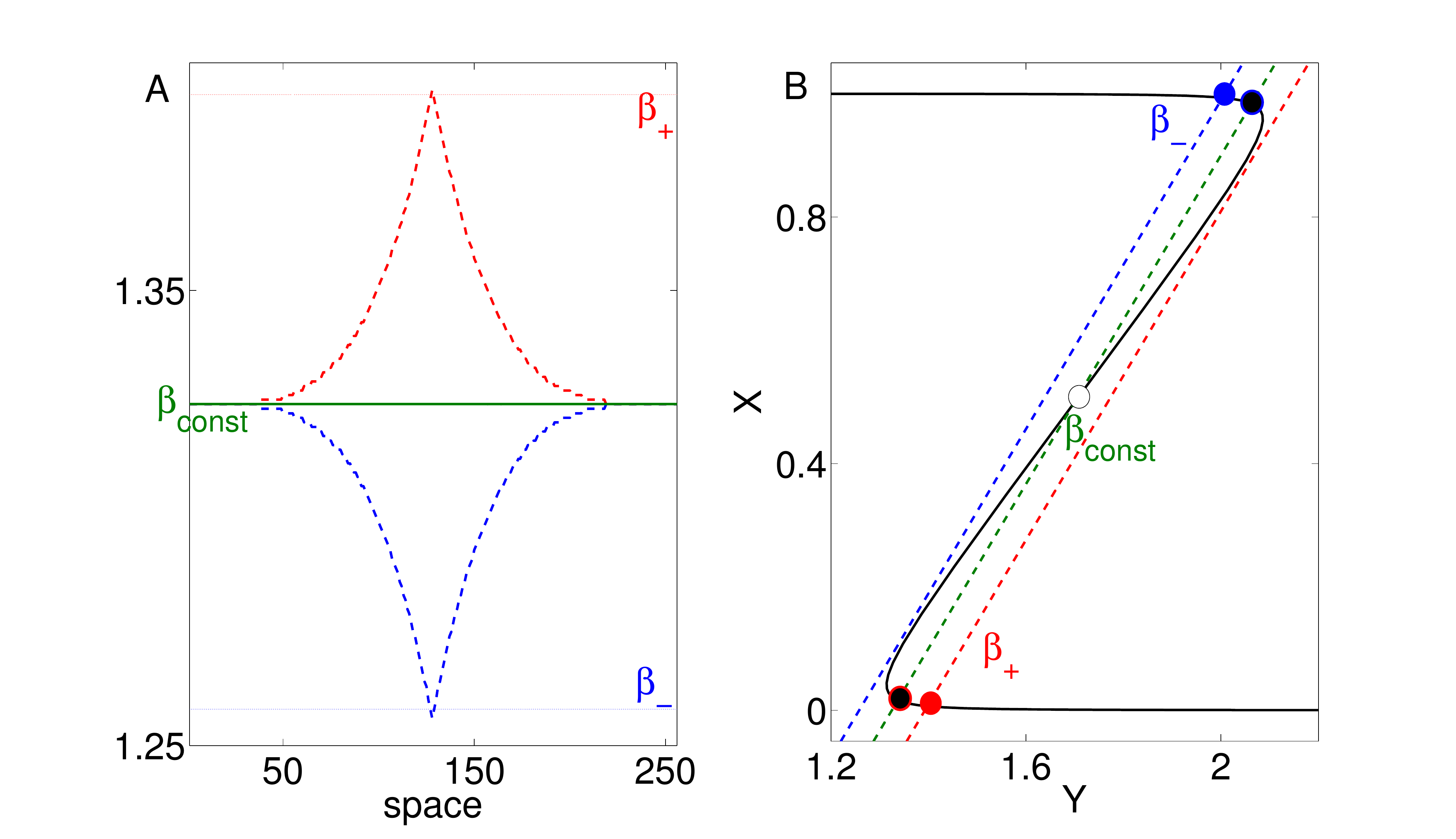} 
\caption{\label{fig4} (Color online) Spatial profiles of $\beta$. A) Uniform (green) and nonuniform $\beta$'s (red and blue). $\beta_+$ and $\beta_-$ are the maximum and minimum values.   B) Intersections of the nullclines for $\beta_+$,  $\beta_{const}=1.325$, and $\beta_-$. Other parameters are the same as in \ref{fig2}.  }
\end{figure}

\subsection{The stability of uniform solutions}
The stability of the uniform solutions, ${\bf X}= (X_0,Y_0)$, can be analyzed by putting the perturbed uniform solutions, ${\bf X}= {\bf X_0}+ \mathrm{e}^{\lambda t} \mathrm{cos(q x) }\delta{\bf X}$,  into the continuous limit of Eq. (4) \cite{bat0}.  After standard calculations, the characteristic equation for the stability of the uniform solutions can be obtained. It is given by,
\begin{equation}
\epsilon_Y \lambda^2+ \epsilon_Y (D_X q^2-f_X +\frac{1+D_Y q^2}{\epsilon_Y})\lambda+ (1+D_Y q^2) (D_X q^2 -f_X)-\gamma f_Y=0,
\end{equation}
where, $f_X=\frac{\sigma^2-f_1^3}{2 f_1^3}$, $f_Y=-\frac{\sigma^2}{2 f_1^3}$, and  $f_1=\sqrt{\sigma^2+(\alpha-X_0 +Y_0)^2}$.  

\begin{figure}
\vspace{-0.1in}
\includegraphics[width=1.\linewidth,height=0.5\linewidth,scale=0.4]{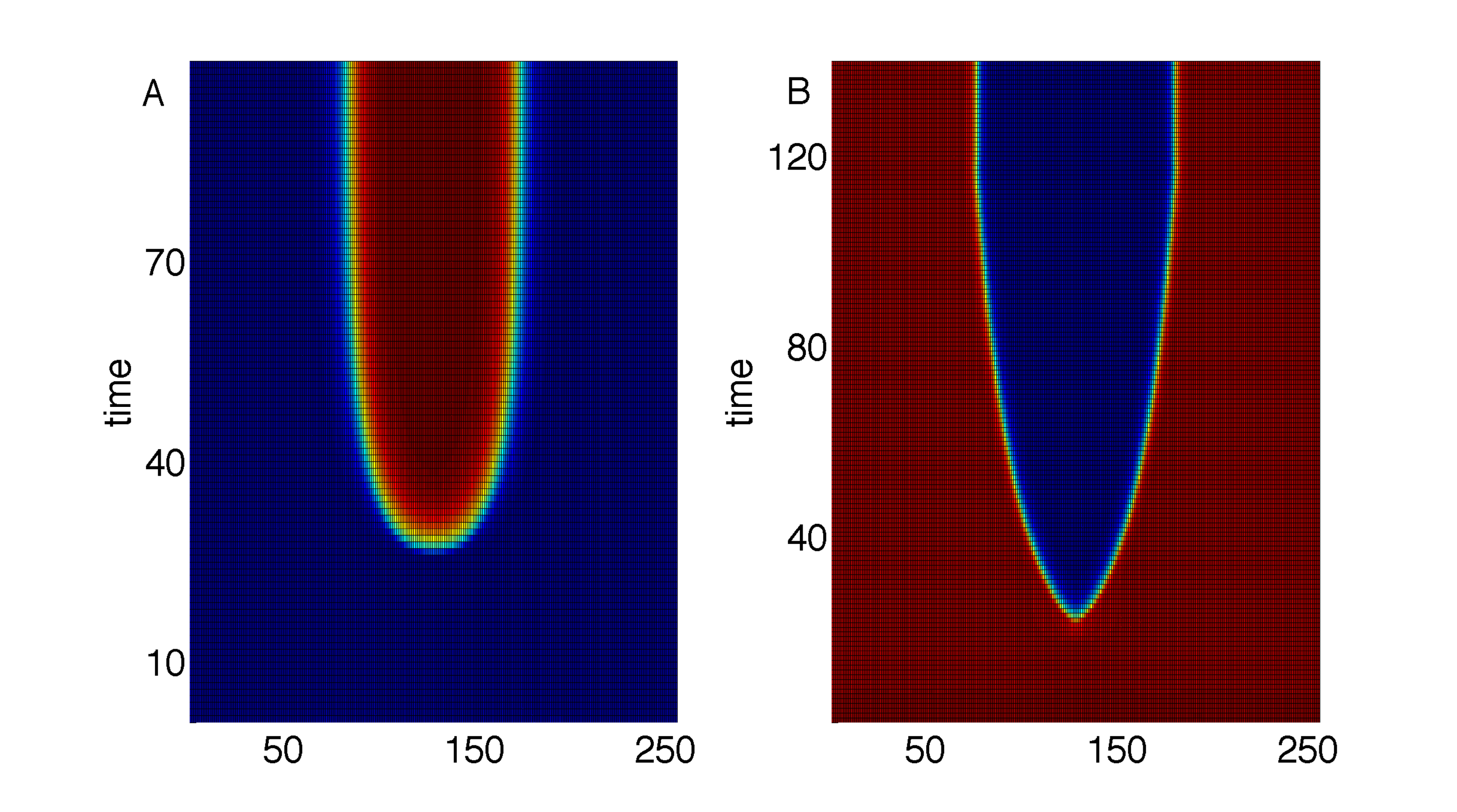} 
\caption{\label{fig5} (Color online) Domain nucleation in Eq. (4) for nonuniform $\beta$'s. A) Domain nucleation from a lower stable steady state of $X$ by nonuniform $\beta$ shown in Fig. \ref{fig4}A by blue dashed lines. B) Domain nucleation from an upper steady state of $X$ by nonuniform $\beta$ shownin Fig. \ref{fig4}A  by red dashed lines.  Horizontal axes plot number of cells. Parameters are:  $D_X=4$, $D_Y=0.4$, and $\epsilon_Y=0.1$. Noflux boundary conditions.  }
\end{figure}

\section{Domain patterns}
\subsection{Domain nucleation by Turing mechanism}
In a certain region of its parameters, Eq. (4) displays a monostable steady state that can undergo Turing instability if $D_Y>>D_X$. The linear stability analysis using Eq. (8)  indicates that the critical wavenumber does not change significantly upon the change of the parameters $\beta$ and $\gamma$.  For instance, in Fig. \ref{fig3} we computed $q_{cr}$ ($\lambda(q_{cr})=\lambda_{max}>0$) at two different values of $\beta$. In constrast to the experimental data \cite{simon} which show a substantial  increase of the size of the central domain  with the down regulation of  CLV expression, the critical wavenumbers in Eq. (8) increase with the reduction of $\beta$ in Fig. \ref{fig3}. Also it can be shown that with the change of $\beta$ (or $\gamma$), the maximal possible change of the critical wavenumber cannot exceed 50\% from its minimal value. Such a limited interval for the critical wavenumbers suggest that Turing instability may have limitations for descring the domain size control in the SAM \cite{simon}. 

\begin{figure}
\vspace{-0.1in}
\includegraphics[width=1.\linewidth,height=0.5\linewidth,scale=0.4]{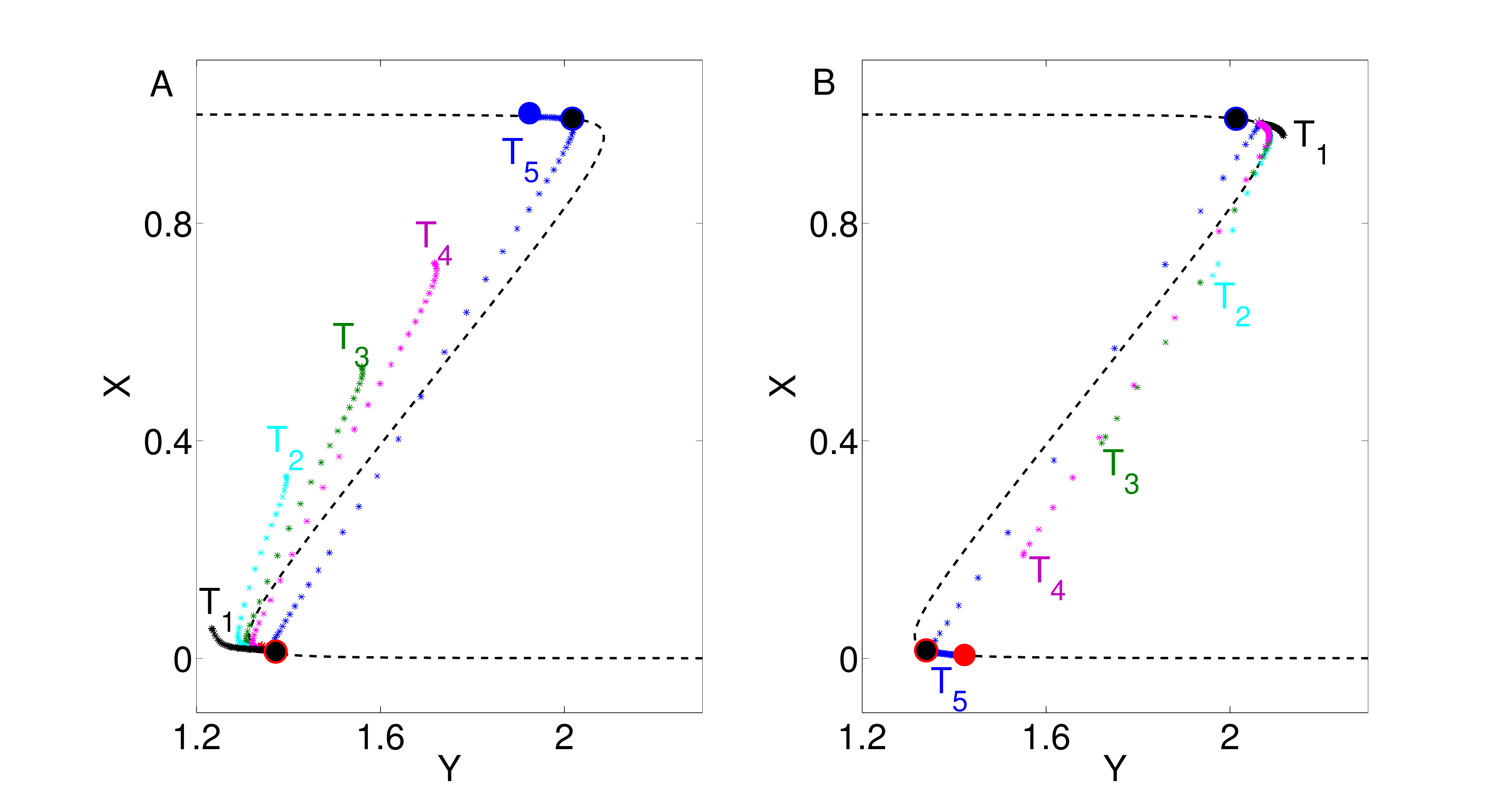} 
\caption{\label{fig6}  (Color online)  Projections of $(X_i(t),Y_i(t))$, $i=1,N$, on the phase plane at different time moments of a domain nucleation. A) Nucleation from a lower steady state. B) Nucleation from an upper steady state.  Parameters are the same as in Fig. 5.}
\end{figure}

Turing instability  is also possible in a bistable regime of Eq. (4), when an upper or lower homogeneous steady-state undergoes Turing instability. A special feature of such an instability occuring only near a saddle-node point is that  a large amplitude pattern can be developed \cite{metens}. The main difference between Turing instabilities in monostable and bistable regimes is that depending on initial conditions, domains of different sizes can emerge in a bistable regime due to the inteplay of Turing patterns and bistability. 

\subsection{Domain nucleation by a bistable switch: nonuniform $\beta$}
Here we consider Eq. (4) in a parameter region where it displays bistability. In Eq. (4), the parameter interval where both homogeneous steady states are stable against Turing instability is much wider than the parameter regions where a monostable or bistable state  is unstable against Turing instability.  In our previous work \cite{bat0},  we studied a model similar to Eq. (1-2), but with a different form of the nonlinear function $\Phi$. We have shown that a nonuniform diffusive field of a peptide-hormone synthesized in the system, can derive domain nucleation  by switching the cells into upper or lower domains. Moreover, such a field can confine the domain at a given location, by stabilizing the fronts connecting the areas with low and high values of $X$.
  
To illustrate the domain nucleation in a two-variable model  Eq. (4), let us assume that each cell has a different value of parameter $\beta_i$.  Also, for the sake of convenience, let us consider the system on $(Y,X)$ plane. As an example, Fig. \ref{fig4}A shows uniform and noniniform distributions of $\beta_i$. If $\beta_i$  is uniform and given by the green line in  Fig. \ref{fig4}A, Eq. (4) displays bistability. For $\beta_i=\beta_{const}$,  depending on initial conditions, two spatially uniform solutions are possible in the simulations of Eq. (4), near the black symbols in Fig. \ref{fig4}B, surrounded by red and blue circles.   Using Eq. (8) it can be shown that these  uniform states are stable against small nonuniform perturbations.

If $\beta_i$ is uniform and given by the red or blue dashed lines in  Fig. \ref{fig4}A, Eq. (4) is monostable; the system evolves  either into an upper state (filled blue circle) or into a lower state (filled red circle),  Fig. \ref{fig4}B. The question is where to the system evolves if $\beta_i$ is nonuniform and given by the dashed lines (blue or red) in Fig. \ref{fig4}A? Intuitively, for the blue dashed lines, if the diffusion coefficients are small and the system is initially in a lower state, the cells  with $\beta_i \approx \beta_{const}$ may remain in a lower state, but the cells with $\beta_i=\beta_-$ may switch to an upper state and remain there. 

\begin{figure}
\vspace{-0.1in}
\includegraphics[width=1.\linewidth,height=0.5\linewidth,scale=0.4]{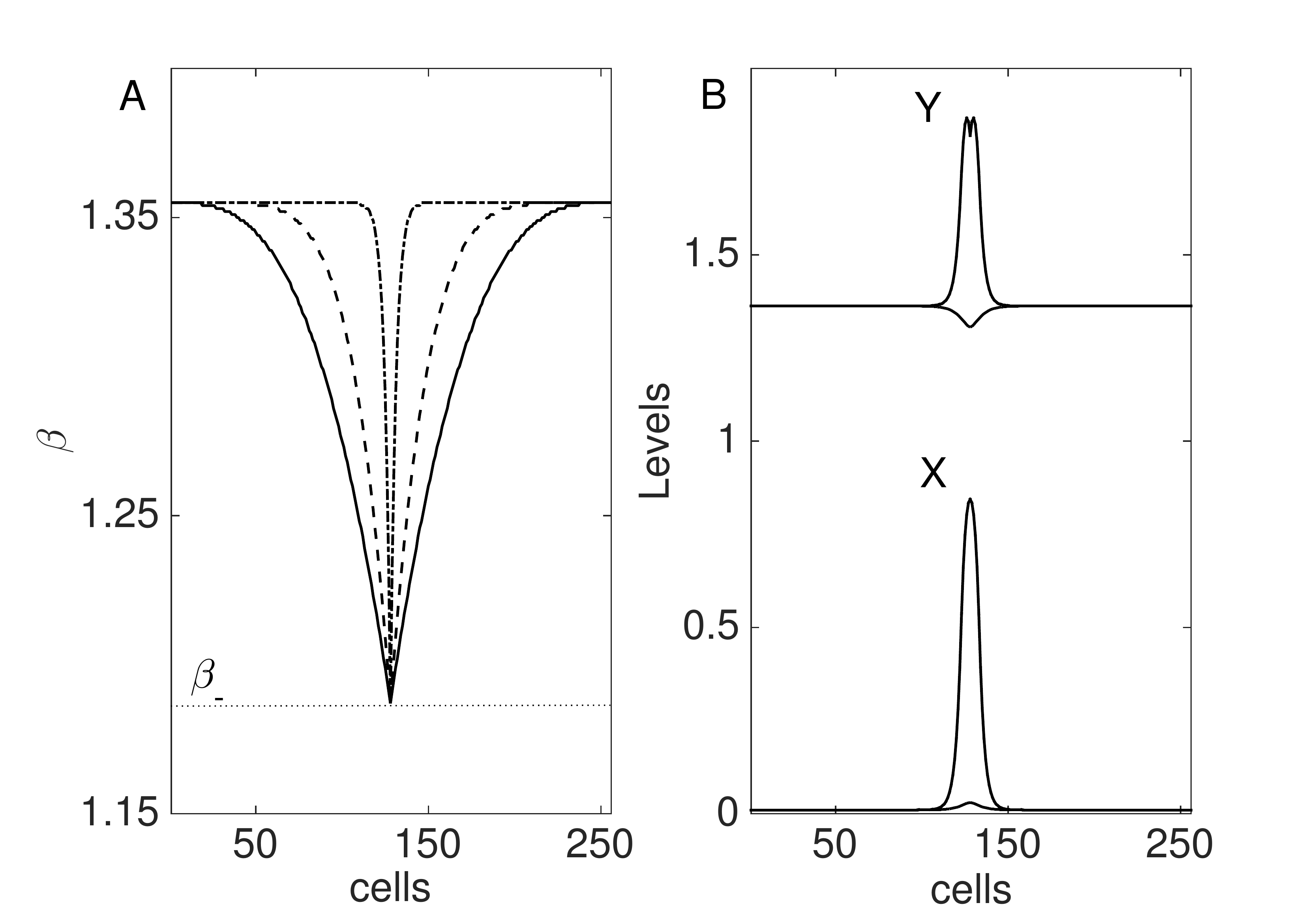} 
\caption{\label{fig4new} Simulations of bistable switches for different $\beta$. A)  Spatial profiles of $\beta$. B) Presence and absence of domain nucleation. Solid curves: $D_X=7.5$ and $D_Y=0$. Dot-dashed curves $D_X=7.5$ and $D_Y=25$. Other parameters are the same as in \ref{fig5}. }
\end{figure}

Fig. \ref{fig5} shows space-time plots of  $X$ in the simulations of Eq. (4) with  $\beta_i$ shown in Fig. \ref{fig4}A by the dashed lines in blue and red. In  Fig. \ref{fig5}, after the nucleation, the domain growth ceases and domains are confined. The reason is that if the fronts connecting upper and lower states is motionless at $\beta_i=\beta_{const}$, the fronts can also be motionless if $\beta_i$'s nonuniform and has a long-wave distribution near  $\beta_{const}$. We refer to Ref. \cite{bat0} for a detailed explanation of domain confinement in a minimal model of SAM. 

Fig. \ref{fig6}  shows the projections of the distributions of $(X_i(t),Y_i(t))$, $i=1,N$,  at different time moments $T_k$, on the $(Y,X)$ plane. Initial conditions are chosen such that the system is either near  the lower, or near the upper steady state. Fig. \ref{fig5}A and  Fig. \ref{fig6}A show that at time $t=T_1$, the levels of $Y$ field at the center of the system  become less than $Y_{SN_2}$.  Thus the elements in the center of the system, where $Y_i(t)<Y_{SN_2}$, become unstable. Consequently, as Fig. \ref{fig6}A shows the unstable elements are switched to the upper domain. Similarly,  Fig. \ref{fig5}B and Fig.  \ref{fig6}B illustrate a domain nucleation from the upper state, for nonuniform $\beta$ shown by the red dashed lines in Fig. \ref{fig4}A, with the maximum value $\beta_+$. 

\begin{figure}
\vspace{-0.1in}
\includegraphics[width=1.\linewidth,height=0.5\linewidth,scale=0.4]{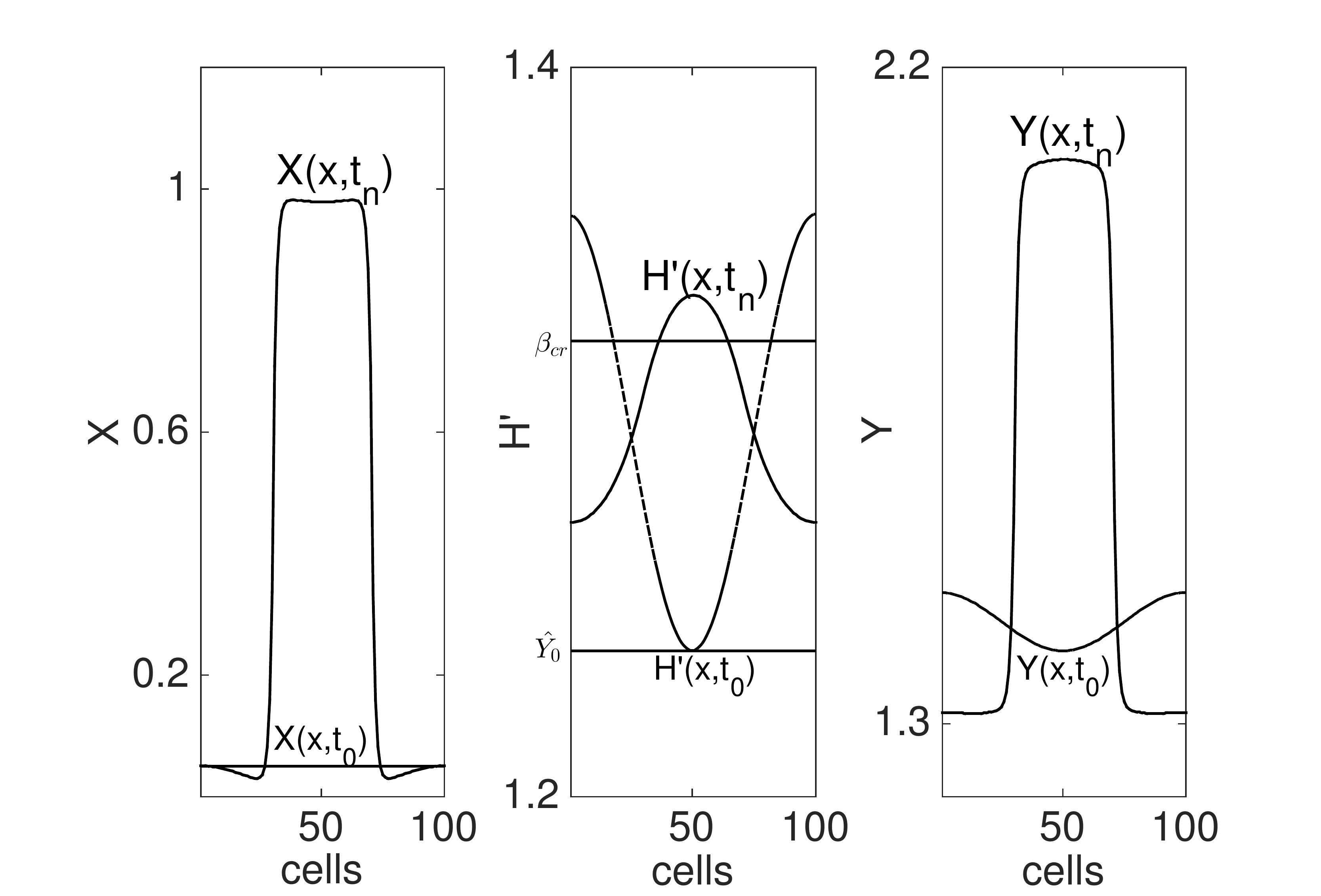} 
\caption{\label{fig8}  Domain nucleation by nonuniform $H$. Distributions of $X$, $H'=\hat Y_0+\mu H$, and $Y$. Distributions at initial time moment, $t=t_0$, and at  $t=t_n$ when the system is in the stationary state. Parameters are: $\gamma=0.75$, $\alpha=1.2$, $D_X=1.2$, $D_Y=0.12$, $D_H=100$, $\sigma^2=0.001$, $\mu=0.15$, $\hat Y_0 =1.24$, $\epsilon_Y=0.1$, $\beta_{cr}=\beta_{const}$, and $\epsilon_H=0.01$.  Noflux boundary conditions.  }
\end{figure}

Our simulations in this subsection show that at the onset of bistability, nonuniformly distributed $\beta_i$ in Eq. (4) can enforce domain nucleation and confinement, through localized switches of  the elements into bistable states.  The occurrence of localized switches is dependent mainly  on the spatial profile of $\beta$, initial distributions of $X$  and $Y$, and  diffusion coefficients $D_X$ and $D_Y$. As an example, we simulated Eq. (4) for different values of $D_X$ and $D_Y$, using different spatial profiles of $\beta$, but with the same $\beta_-$. The initial distributions of $X$ and $Y$ were set near the lower state corresponding to the state at  $\beta_i \approx 1.35$. When $\beta$ has a profile as shown in Fig. \ref{fig4new}A by the dot-dashed lines, no elements were switched from the initial lower state to a higher state, for sufficiently large $D_X$ and $D_Y$. However, if $D_Y = 0$, the elements are forced to switch, Fig. \ref{fig4new}B solid lines. Therefore, in contrast to Turing instability, which requires $D_Y>>D_X$ for pattern formation to be possible, a domain nucleation is possible  by the bistable switch when $D_X>D_Y=0$. 

\subsection{Domain nucleation by a bistable switch: nonuniform $H$ field in Eq. (1-2)}
The results of the previous subsection,  i. e. domain nucleation in Eq. (4) for spatially nonuniform $\beta$, suggest that domain nucleation can be possible in  Eq. (1-2)  if the $H$ field has a nonuniform, long-wave distribution. Moreover, if $H(x,t) $ is distributed around $\beta_{const}$, the nucleated domain can be stationary.  

Let us choose the parameters such that the reaction model of Eq. (1-2) displays a monostable, upper fixed point, and such that  the uniform solutions are stable against Turing instability. Fig. \ref{fig8} shows simulations of Eq. (1-2) from a nonuniform initial distribution of $H$. The spatial distributions of the state variables are shown at two different time moments: initial and  final time of the simulations when the pattern became stationary. A simple  explanation of domain nucleation is as follows. Because uniform solutions near a lower state do not exist, $X$ and $Y$ fields are attracted to the upper fixed point. However, because the system is near the  onset of bistability, a nonuniform, stationary distribution of $H$ is possible, by the bistabile switch in the system. Fig. \ref{fig8}B shows that the stationary distribution of the variable $H'=\hat Y_0+\mu H$ is spatially nonuniform and distributed near $\beta_{cr}=\beta_{const}$. 

\begin{figure}
\vspace{-0.1in}
\includegraphics[width=1.\linewidth,height=0.5\linewidth,scale=0.4]{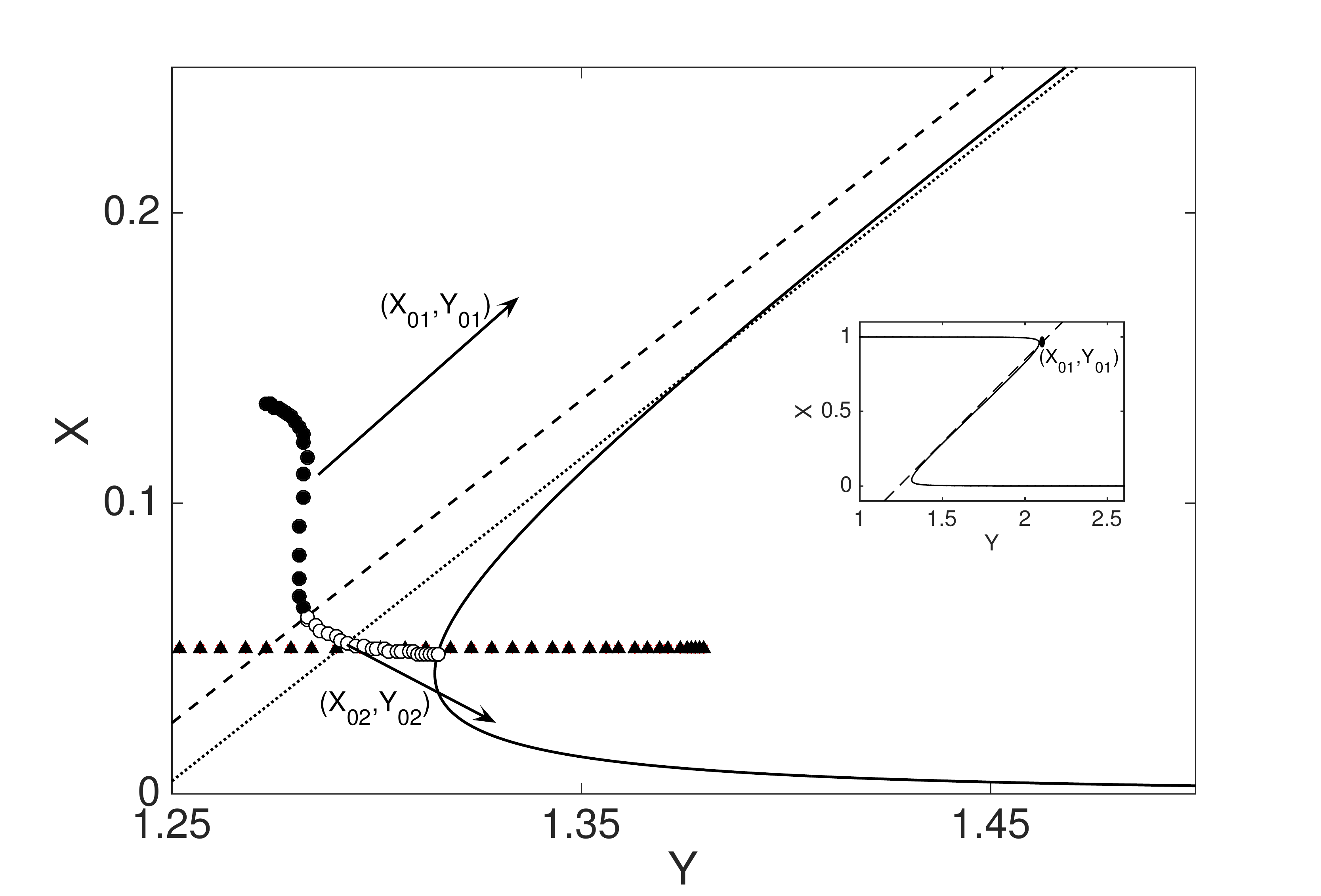} 
\caption{\label{fig9} A phase plane view of a domain nucleation by nonuniform $H$. Symbols show $X$ and $Y$ distributions at a given time moment. Solid lines show $X$ nullcline and dashed lines show $Y$ nullcline for the homogeneous system. Dotted lines indicate that the system is entering into a bistable regime at the onset of domain nucleation, when the distributions acquire a profile shown by the open and filled circles. The inset shows the homogeneous steady state.  Parameters are the same as in Fig. \ref{fig8}. }
\end{figure}

Domain formation by bistable switch strongly depends on the spatial distributions of the state variables. Therefore, it is desirable  to be able to discriminate initial distributions of $X$, $Y$, and $H$ fields that lead to stable domain patterns. For example, in  Fig. \ref{fig9} we illustrate an onset of domain nucleation on $(Y,X)$ plane. If the initial distributions are all uniform, the system will evolve into the fixed point shown in the inset of  Fig. \ref{fig9}, given by the intersections of $Y$ (dashed lines) and $X$ (solid lines) nullclines. The triangles show a nonuniform initial distribution along the $Y$ axis, near a lower steady state,  which evolves further into the distribution shown by small circles at the onset of a domain nucleation. We computed an effective intercept at the time moment of domain nucleation, in the global coupling limit ($D_H>>1$), $\beta_{eff} \approx \hat Y_0+\mu \overline{X(x,t)}$, where the bar represents spatial average of $X$ at the time of approximation. If we overplot $Y_{eff}(t)=\beta_{eff}+\gamma X$ in Fig. \ref{fig9}, we obtain the dotted lines in Fig. \ref{fig9}, which suggest that with the  effective $\beta_{eff}$, the system is at the onset of emergent bistability. To which of the fixed points local cells will be attracted depends on the effective spatial distributions. The cells around the peak of the distribution are driven by $H$ with a stronger intensity,  and they are attracted to the upper fixed point, Fig. \ref{fig9} filled circles; whereas, the cells near the bottom of the distribution are driven by $H$ with a weaker intensity, and they are attracted to the lower fixed point, Fig. \ref{fig9} open circles. 

Our analysis and simulations show that if the initial distributions are uniform or near uniform, $H$ field modulates $\gamma$, the slope of $Y$ nullcline. Then no domain nucleation is expected. However, if the  distributions are nonuniform, $H$ field modulates the parameter $\beta$ and domain nucleation is possible, because the system behaves as if $\beta$ has a nonuniform distribution.  In other words,  $H$ field acts as a bifurcation parameter; depending  on initial distributions, $H$ field can separate the elements into bistable domains. The size of a nucleated domain depends on the parameters of $\beta$ and $\mu$, which control the front velocity \cite{bat0}. Typically, smaller is $\mu$, larger is the size of a nucleated domain \cite{bat0}. We plan to study in a separate work to classify the initial conditions for $H$ that lead to the emergent bistability and robust domain patterns in Eq. (1-2).

\subsection{Subsequent domain nucleations in  growing systems}
As plants grow new domains of high WUS levels nucleate in the SAM. Mathematical models of SAM should be able to describe  domain nucleations as the size of the system increases. 
\subsubsection{Turing Instability}
In a monostable system of  a fixed size $L$, where a homogeneous steady state is undergoing Turing instability,   the number of nucleated domains can be estimated by $n_{dom}=\frac{2\pi L}{q_{cr}}$. $n_{dom}$ is eventually the same if the size of the system was smaller than $L$ initially and the initial number of the nucleated domains was less than $n_{dom}$. In a cell network, growing by cell growth and divisions, $n_{dom}$ is also defined by the critical wavenumber \cite{fujita}.

In growing bistabe systems, $n_{dom}$ is dependent on initial conditions and a growth rate. Pattern formation by Turing mechanism in bistable systems can generate domains of different sizes. Namely, if the growth of the system is slow, subsequent domain nucleatios may generate domains of larger sizes; whereas, if the growth of the system is fast, domains can form a periodic pattern.  

\subsubsection{Bistable switch}
Fig. \ref{fig8} shows that after the nucleation of the central domain, the system is effectively bistable, despite the homogeneous reaction system is monostable. The parameters have been chosen such that the uniform solutions at both upper and lower steady states of the effective bistability are stable against Turing instability. However, subsequent, autonomous, domain nucleation is possible in Fig. \ref{fig8}  if the size of the system starts to increase. In other words, the system can enter  back to the  monostable regime locally in the areas further away from the central domain. As an example, consider a case when new elements $X_{new}$ and $Y_{new}$ are added at the boundary at time $t$, as a result of a growth process, with the same values as the boundary elements $X_{0,N}$ and $Y_{0,N}$. Because of  higher $H$ in the center of the system, the system remains bistable there. However, because the fronts are motionless, the local $H$ values near the  boundaries can become low and reach the value $\hat Y_0$ shown in the middle plot of Fig. \ref{fig8}, at which the reaction system is monostable. Therefore, monostable and bistable regimes coexist in the system temporarily, until new upper domains are nucleated near the boundary, restoring the effective bistability globally. Thus in growing systems, new domain are nucleated by the bistable switch mechanism in the areas where the fast diffusive field drops below the critical level. For more detailed description of subsequent domain nucleation we refer to Ref. \cite{bat0}.

\section{Domain patterns in a cell network: Turing vs agent driven instability}
From a mathematical modeling point of view, SAM is a complex multiscale system; therefore, a realistic model of SAM should account for reaction and diffusion processes in individual cells, cell-to-cell interactions, and the dynamics of the cell network due to growth and cell division. In our current models,  we do not take into account cell network dynamics; instead, we adopt a traditional approach in the theory of weakly coupled systems, by considering the cells as identical entities with well defined dynamics, which are under a weak force of mutual interactions  \cite{asm,kurabook}. In this approach, the individual dynamics of the elements do matter; depending on whether the intrinsic dynamics is monostable or bistable, the system evolves into qualitatively different states. 

In this section, we compare the two mechanisms of domain nucleation in Eq. (1-2) and Eq. (4): Turing instability in a monostable system and the agent controlled switch in a bistable system, on a  cell network composed of $N$ polygonal cells, representing a longitudinal cross-section of SAM. Note that the diffusion terms in Eq. (1-2) and Eq. (4) should be replaced by $\mathit{D_{iff}} Z_i=D_Z $ $\sum_{i=neighbors}^{N} (Z_{j}-Z_{i}) $, $Z=X,Y$, to adjust to the number of neighbors in  the network geometry.  Following Ref. \cite{nikolaev}, we assumed in our simulations  that $Y$ (CLV) expresses strongly along the first cell layer.  We note the results of this section are qualitatively the same if we dismiss this assumption and use just no-flux boundary conditions.  Also, the results are qualitatively the same, if we modify the form of a cell network and cell distributions in it. An algorithm for generation of cell network is given in Ref. \cite{nikolaev}. We solve the reaction diffusion system using finite difference scheme at the center of cells. Assuming well mixing inside the cells, we assign to each cell the color coding corresponding to the WUS or CLV levwls at center of the cells. 

\subsection{Turing instability in a cell network described by Eq. (4)}

\begin{figure}
\vspace{-0.1in}
\includegraphics[width=0.9\linewidth,height=0.45\linewidth,scale=0.4]{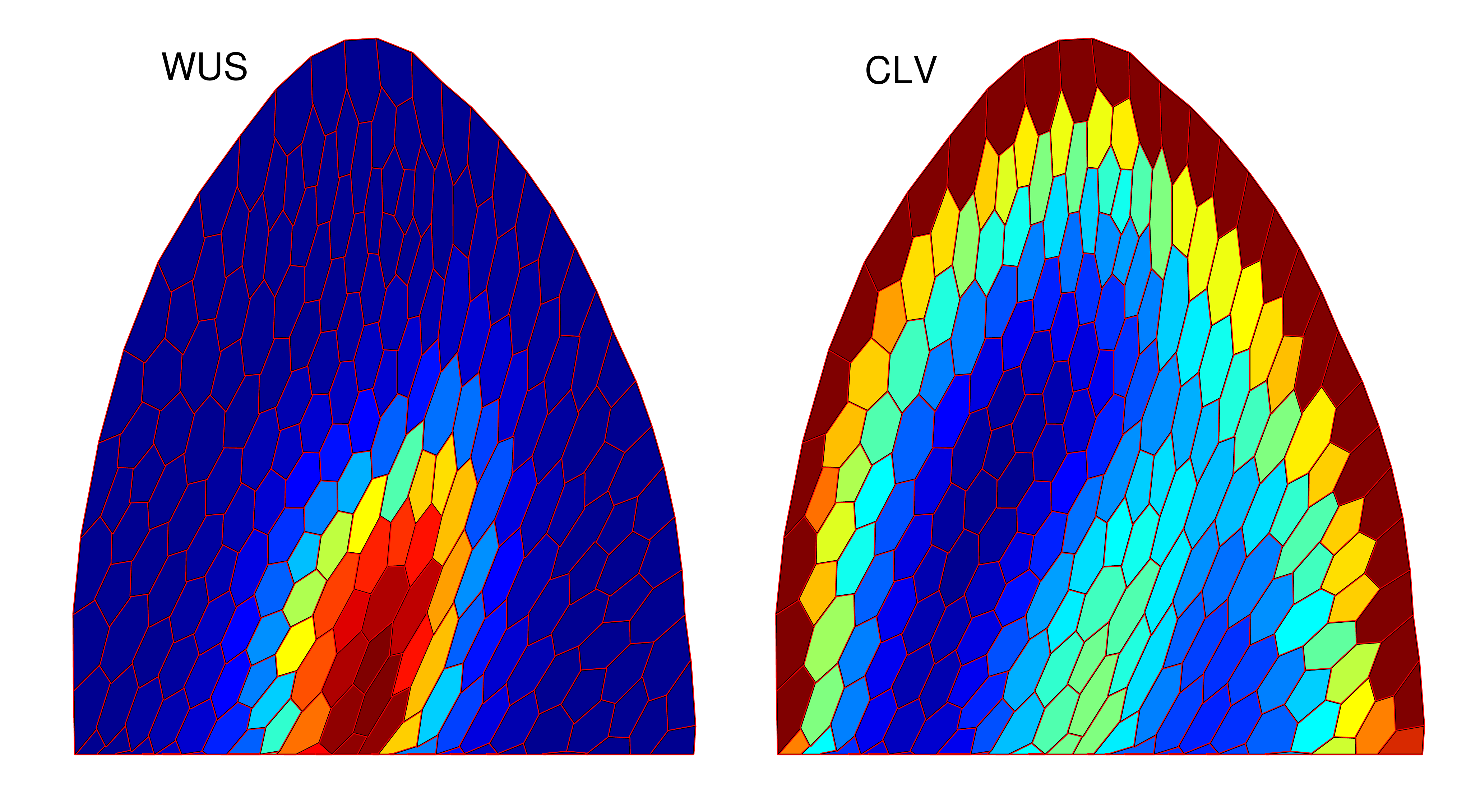} 
\caption{\label{fig11}   (Color online) A stationary domain in the simulations of Eq. (4) on a cell network.  Initial conditions are small random perturbations of the monostable steady  state. CLV is set to $Y_{bnd}=1.8$ along the first cell layer. No-flux boundary conditions are on the bottom. The length of the cell network is $L \approx 8.5$, in both horizontal and vertical directions. Parameters are the same as in Fig. \ref{fig3}.  }
\end{figure}

Fig. \ref{fig11} shows an example of domain formation by Turing mechanism. In our simulations, for initial conditions as random perturbations of the uniform solutions, a domain pattern appears on a cell network at a random location. The same result was  reported for modeling domain formation by Turing mechanism in a detailed model of SAM simulated on a cell network representing a longitudinal cross-section of SAM \cite{simon}. To simulate domain nucleation in the location known from SAM experiments \cite{laux}, i.e. around the fourth cell layer from the apex, and to retain it at this target position, {\em anchoring} was introduced in Ref. \cite{simon}. 

\begin{figure}
\vspace{-0.1in}
\includegraphics[width=0.7\linewidth,height=0.45\linewidth,scale=0.45]{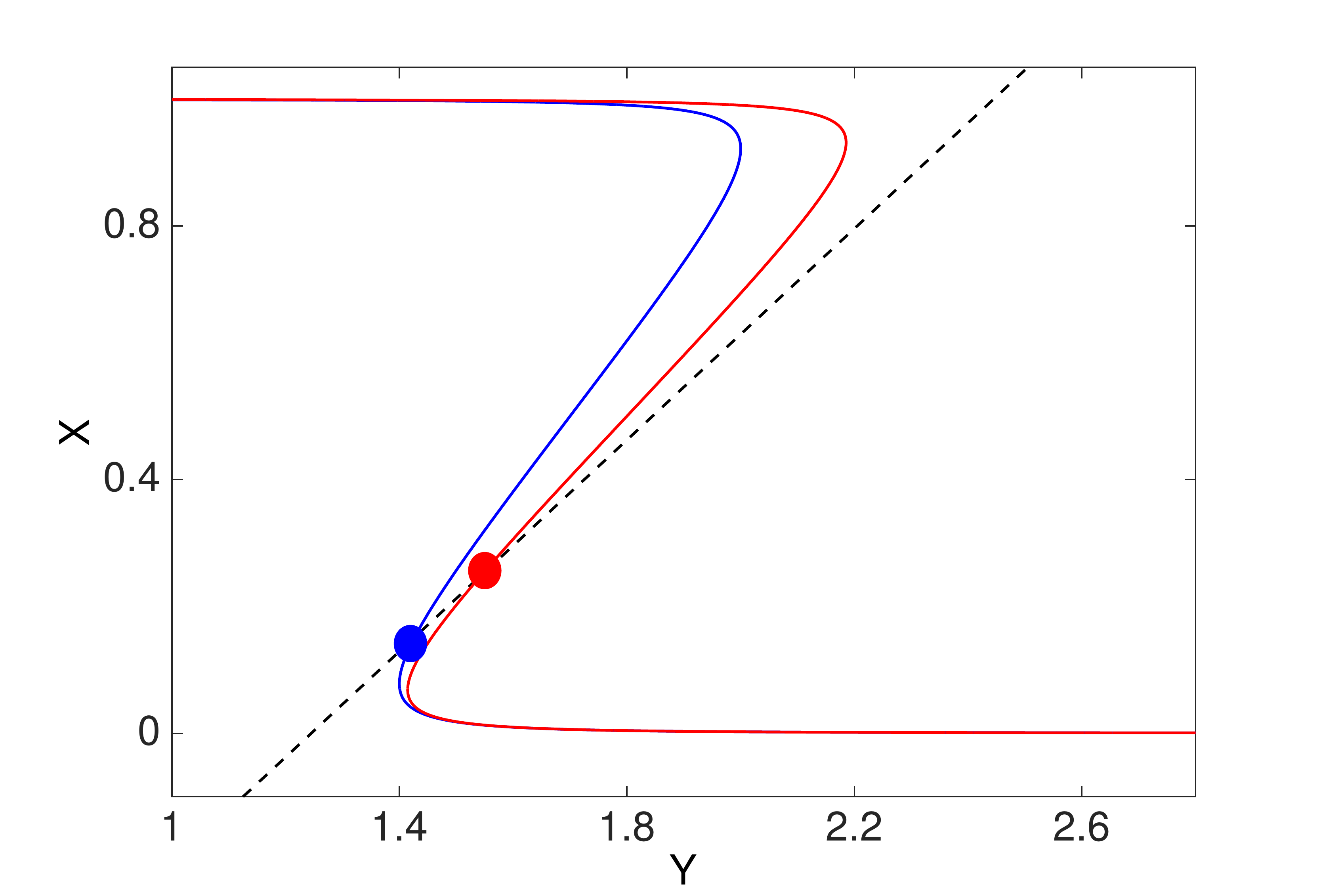} 
\caption{\label{fig12} (Color online) Introduction of a heterogenous parameter in the system. $\kappa_i$ of the nonlinear function $\Phi_{\sigma}(\alpha+\kappa_i X -Y)$ is a step function. Red nullcline $\kappa_i=1$, and blue nullcline  $\kappa_i=1.2$. Other parameters are the same as in Fig. \ref{fig2}. }
\end{figure}

{\em Anchoring} assigns different rates of WUS synthesis in different SAM zones. In  Eq. (4), anchoring can be introduced by modifying the argument of the nonlinear function, by introducing a parameter $\kappa_i$ in the nonlinear function $\Phi_{\sigma}(\alpha+\kappa_i X_i -Y_i)$, as a step function in  space. For instance,  $\kappa_i=\kappa_0$ in the central zone, while $\kappa_i=\kappa_1$ in other zones of SAM.  Fig. \ref{fig12} shows the intersections of $X$ and $Y$ nullclines at $\kappa_i=1$ and  $\kappa_i=1.2$, indicating that the reaction system is heterogeneous if $\kappa_i$ is a step function.

\begin{figure}
\vspace{-0.1in}
\includegraphics[width=0.9\linewidth,height=0.45\linewidth,scale=0.4]{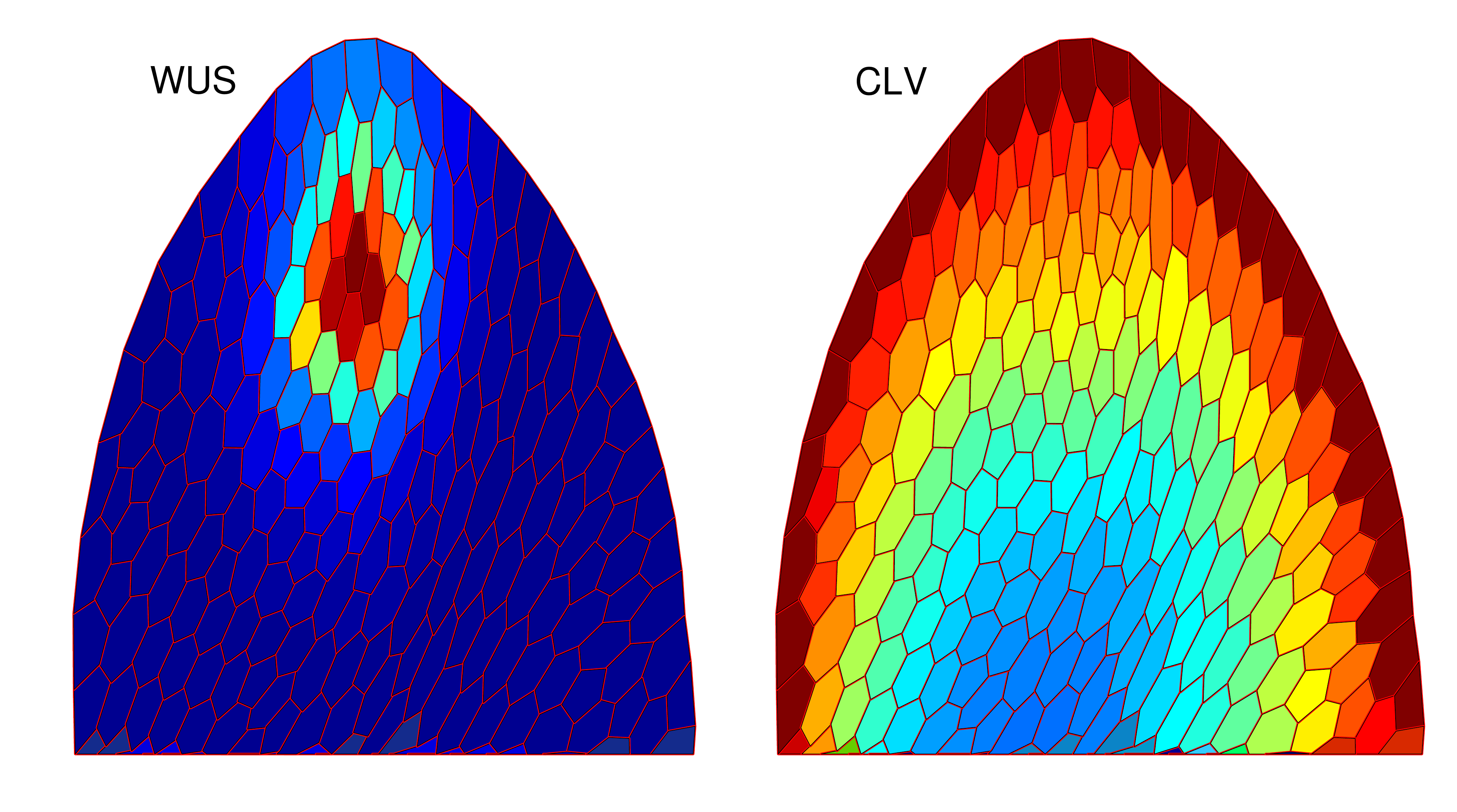} 
\caption{\label{fig13} (Color online) Fixing central domain location by anchoring. Parameters are the same as in Fig. \ref{fig11}.  }
\end{figure}

Anchoring allows domain nucleation to occur at the target location, independent on initial conditions, Fig. \ref{fig13}. Obviously, if domain nucleation involves anchoring, the mechanism is no longer the pure Turing instability, because
the domain size is not defined by the critical wavenumber. Here the actual domain size is defined by the size of the region where the parameter is heterogenious. A steep, nonuniform in space $\kappa_i$ enforces a dynamic heterogeneity in the system, independent of the critical wavenumber selection process. Although such a heterogeneity can be easliy introduced in a mathematical model, an existence of a molecule enforcing such spatial  heterogeneity in SAM needs to be established.

\subsection{Bistable switch in a cell network described by Eqs. (1-2)}
\begin{figure}
\vspace{-0.1in}
\includegraphics[width=0.9\linewidth,height=0.45\linewidth,scale=0.4]{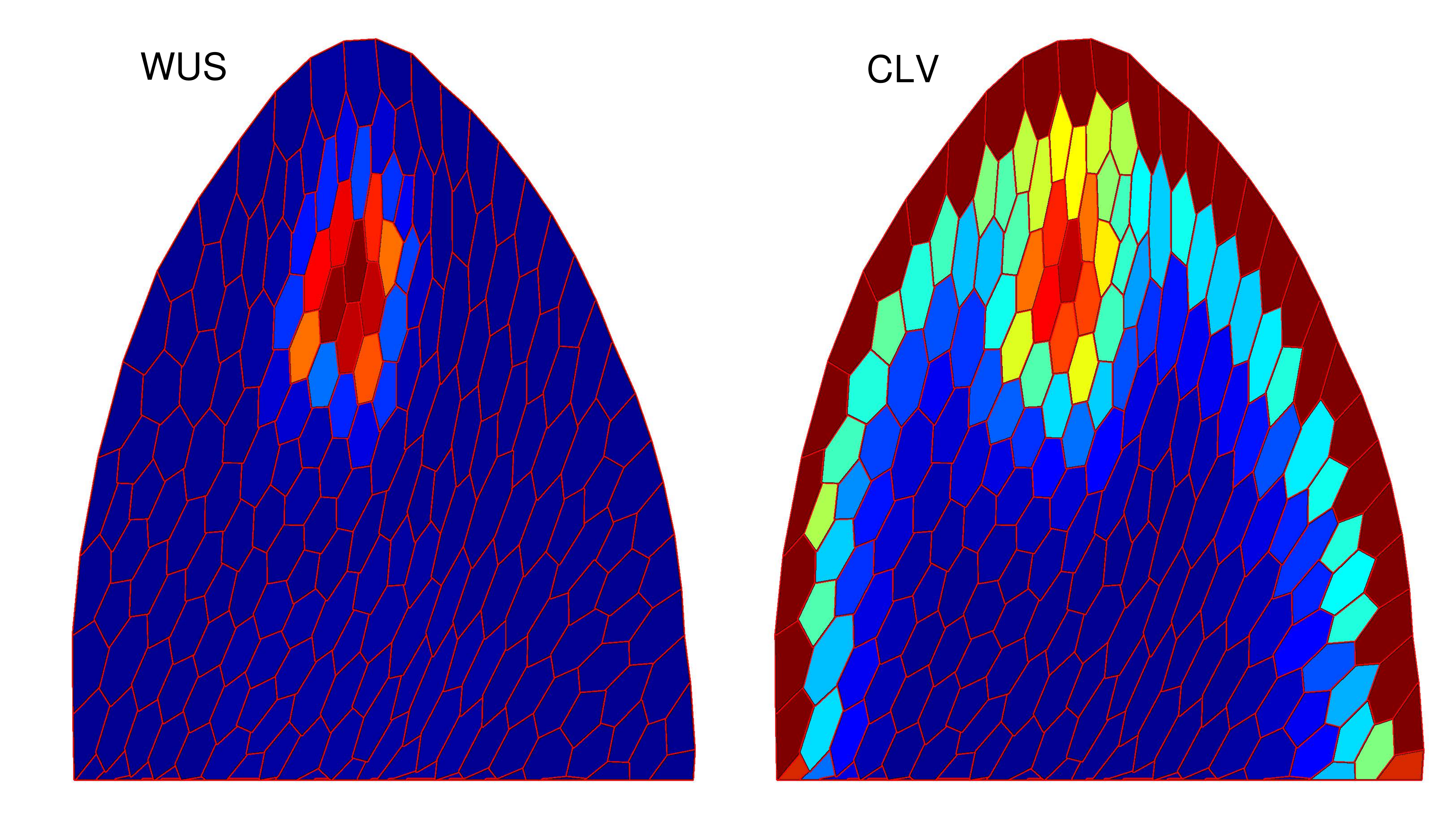} 
\caption{\label{fig14} (Color online)  Domain nucleation at the target location, from initial conditions near upper( for central domain) and lower steady states( for other SAM zones). Parameters are the same as in Fig. \ref{fig9}.  }
\end{figure}

The additional  variable $H$ and the parameters associated with it provide extra means for controlling domain nucleation in Eqs. (1-2), compared to Eq. (4).  Initial distributions of the variables on the cell network, including boundary conditions of $H$ field, can determine the location of domain nucleation the same way as in Fig. \ref{fig8}, the case of one dimensional cell arrays. The size and location of a domain can be controlled by the parameters controlling the diffusive field $H$ \cite{bat0}. Because the effective system is bistable, the domain is stationary. Thus with the bistable switch mechanism, for appropriate initial conditions, a stationary domain can be nucleated without introducing {\em anchoring}, at the target location on a cell network with complex boundaries, Fig. \ref{fig14}.  

\begin{figure}
\vspace{-0.1in}
\includegraphics[width=0.9\linewidth,height=0.45\linewidth,scale=0.4]{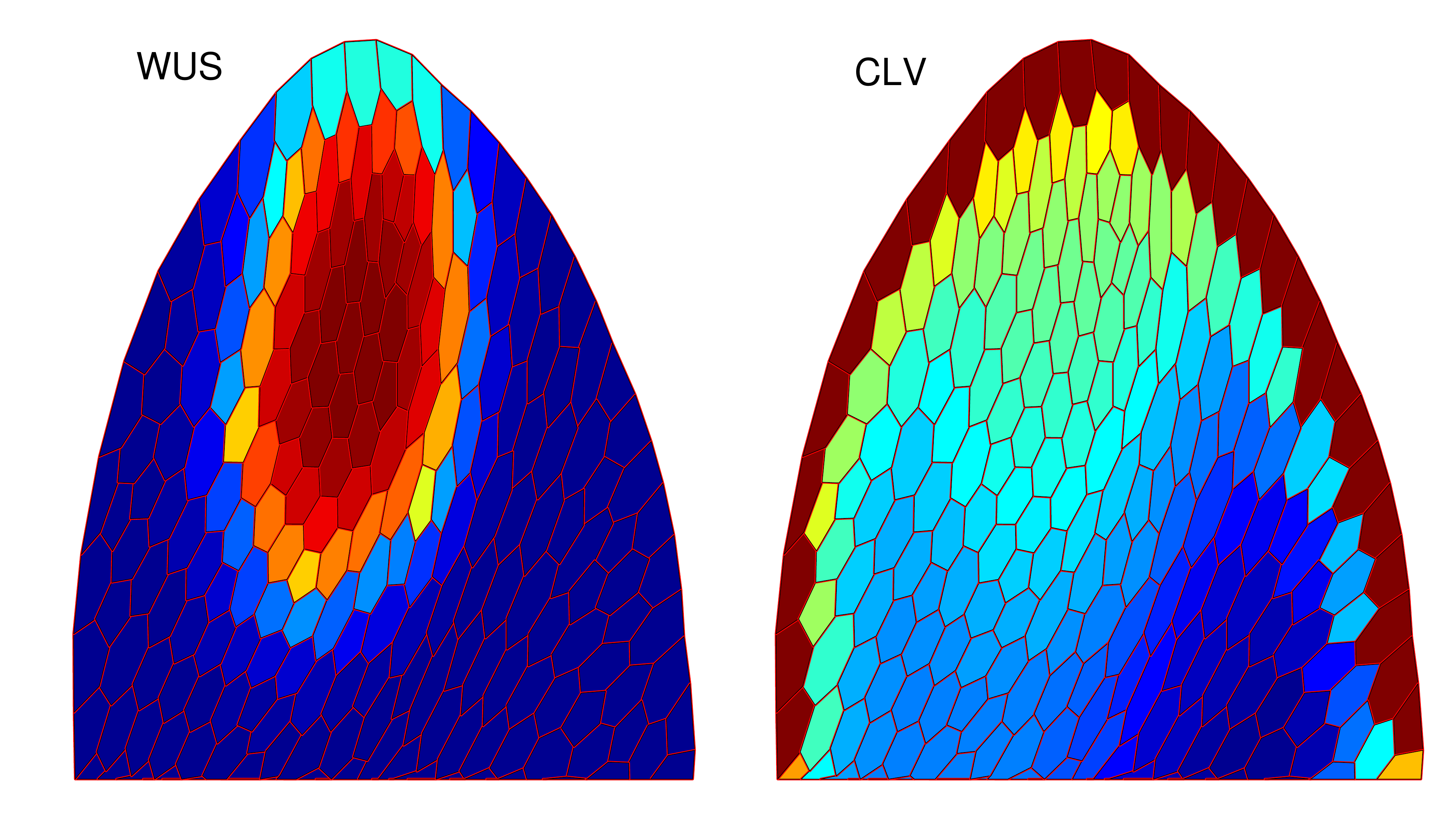} 
\caption{\label{fig15} (Color online)  Domain size increases with the down regulation of CLV. Parameters are the same as in Fig. \ref{fig9}.  }
\end{figure}

It is well known that when the expression of CLV is repressed, the size of the central domain is enlarged \cite{brand,schoof}. Existing mathematical models based on Turing mechanism explain this experimental observation by the decrease of the critical wavenumber of Turing instability, upon the reduction of the parameters controlling CLV expression \cite{fujita}. However, as we have shown in Fig. \ref{fig3}, the change of $q_{cr}$ can be insignificant in Eq. (4), for domain patterns nucleated by the Turing mechanism.  In  Fig. \ref{fig15} we simulate the domain size increase by the bistable switch mechanism. Compared to the simulations in Fig. \ref{fig14}, the expression level of CLV is lowered in  Fig. \ref{fig15},  by reducing the parameter values of  $\hat Y_0$ in Eq. (1-2). As a result, the size of the central domain in Fig. \ref{fig15} is larger than the  size of the central domain in Fig. \ref{fig14}. Also, WUS expression is extended up to the first cell layer in Fig. \ref{fig15}. Mathematically, the enlargement of the domain size can be explained by the dynamics of the front connecting the areas with high and low $X$ levels. We refer to Ref. \cite{bat0}, for the details of front dynamics in a agent controlled system.

\begin{figure}
\vspace{-0.in}
\includegraphics[width=0.9\linewidth,height=0.45\linewidth,scale=0.4]{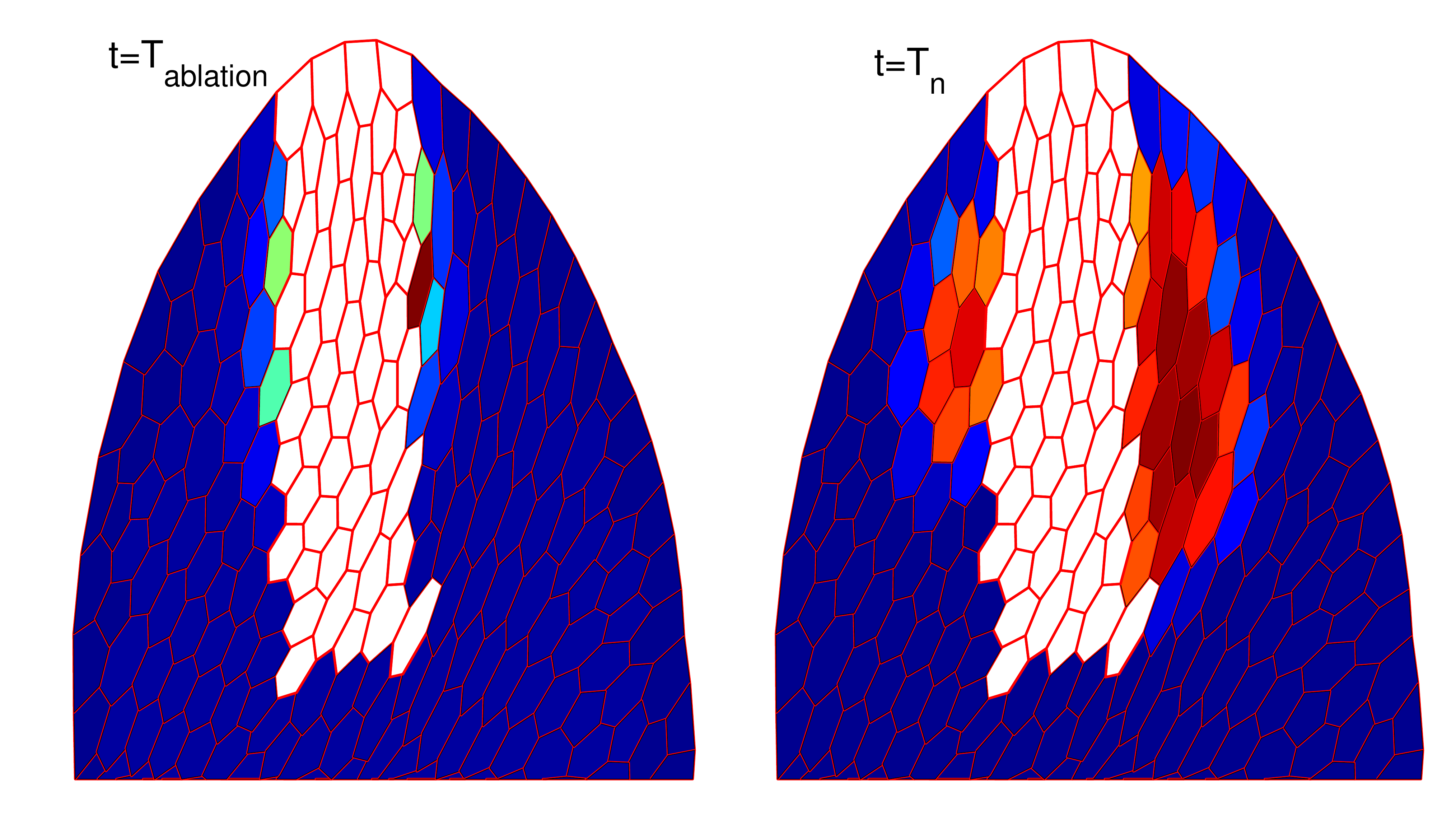} 
\caption{\label{fig16} (Color online)  Simulation of laser ablation experiment. Left figure shows distribution of WUS after the moment of laser ablation. Right figure shows nucleation of two new domains at a later time moment.  Parameters are the same as  in Fig. \ref{fig9}.}
\end{figure}

Next we simulate the laser ablation experiment \cite{reinhardt}. It was shown that if the cells in the central domain are treated with laser, they  no longer express genes and the central domain ceases to exist. However, it was observed that the cells surrounding the dead central domain start to express WUS  \cite{reinhardt}.   In our simulations in  Fig. \ref{fig16}A, the cells shown in white represent the laser treated cells. Because of the WUS depletion in the treated cells, the level of $H$ drops in the cells neighboring the treated cells. Mathematically, after the laser treatment, the system reenters into the monostable regime, see Fig. \ref{fig8}. As a result, a new domain formation by the bistable switch is possible. Fig. \ref{fig16}B shows the formation of two new domains of WUS expression.  Depending on initial conditions and parameter values, nucleation of a single doman can also be simulated \cite{reinhardt}. In earlier works, domain formation following laser ablation have been explained by Turing mechanism \cite{simon,fujita}. The main difference between Turing and bistable switch mechanisms is that in the latter case there is no predetermined critical wavenumber that should be adjusted to the geometry of the system. In other words,  in the bistable switch mechanism, domain nucleation is controlled by $H$'s level and it can be initiated anywhere where its level can become low, for example, by modulating $H$'s level at boundaries.

\section{Discussion}
We have shown that at the onset of a saddle-node bifurcation, a nonuniform field of a fast, diffusive agent can derive domain nucleation, by switching weakly coupled elements into different states. The stationary distribution of the nonuniform field is dependent on initial conditions;  whereas, into which state the elements will be switched to is dependent on the spatial profile of the fast diffusive field. Typically, if the forcing field is near to its maximum, it switches the elements into the higher domain, but if the forcing field is near to its minimum, it switches the elements into the lower domain. If there exists a standing front solution in the system for an uniform field, the nucleated domain is confined at a given location, even when the field has a nonuniform, long wave distribution around the uniform field \cite{bat0}.

We compared two different autonomous  mechanisms of domain nucleation: Turing instability and agent controlled bistable-switch. There are a few differences between the two mechanisms. First, Turing mechanism is an instability of a monostable system; whereas,  bistable switch is possible only at the interface of mono and bistability. The uniform solutions can be stable against Turing instability for the bistable switch mechanism. Second, there is a critical wavenumber for Turing instabilty, leading to a periodic pattern, independent of initial and boundary conditions; whereas, non-periodic complex patterns are possible for the bistable switch depending on initial and boundary conditions. Third, the size of a domain pattern is determined by the critical wavenumber; whereas, the size of a domain is determined by the condition of standing front solutions. Fourth, for  the existence of stationary  Turing patterns, the diffusion coefficient of an inhibitor should significantly exceed the diffusion coefficient of an activator; whereas, this restriction is not required for the bistable switch. Because the diffusion coefficients of WUS and CLV complex are yet unknown, with the possibility that both can be slowly diffusing molecules, domain formation by Turing mechanism requires validation by detailed measurements of the diffusion coefficients of the active variables in the SAM zones.  

As we have shown in this work, a limitation of the pattern formation by Turing mechanism \cite{simon} is that  an additional {\em anchoring}  assumption is required for targeted positioning  of a domain on a cell-network of complex geometry. We argue that if pattern formation involves {\em anchoring} - an inhomogeneous distribution of a parameter, it can be considered as an enforcement of  a localized heterogeneous distribution in the system, acting independently from the critical wavenumber selection.  In contrary, domain nucleation at a target position by a bistable switch does not require an {\em anchoring} assumption, because dynamic heterogeneity is intrinsic for a bistable reaction system.  

In this work we assumed that the cell network is stationary, however, a realistic model  of SAM involves a dynamic network undergoing pattern formation instability \cite{fujita}. It is interesting to study the interplay between a bistable model of SAM and cell-network dynamics, to establish the conditions of bistable switches and standing fronts on a dynamic cell-network. A cell-network model, where the cell growth and division are described by a minimal model of cell cycle  \cite{batprl}, can be integrated to a minimal model of SAM.  

A detailed quantitave model of stem cell regulation in SAM can significantly advance the knowledge about how stem cell maintenance and proliferation can be controlled. In its turn, the progress in the SAM research can be helpful in increasing the crop production for the increased food demand. We believe that a multiscale complex model with well defined underlying mechanisms will lead to the next step in mathematical modeling of stem cells in SAM, from qualitative descriptions of  observed patterns to predictive phenotypes of genetic and  hormonal controls.

\bibliography{basename of .bib file}

\end{document}